\documentclass[preprint]{elsarticle}

\usepackage{dcolumn}
\usepackage{amsmath}
\usepackage{graphicx}
\usepackage{epsf}
\usepackage{psfrag}
\usepackage{epsfig}
\usepackage{amsfonts}
\usepackage{slashed}
\usepackage{epstopdf}

\newcommand{\bq}{\begin{eqnarray}}
\newcommand{\eq}{\end{eqnarray}}
\newcommand{\be}{\begin{equation}}
\newcommand{\ee}{\end{equation}}
\newcommand{\bea}{\begin{eqnarray}}
\newcommand{\eea}{\end{eqnarray}}

\begin{document}

\begin{frontmatter}

\title{Guises and Disguises of Quadratic Divergences}

\author[UFMG]{A. L. Cherchiglia} 
\ead{adriano@fisica.ufmg.br}
\author[UFMG]{A. R. Vieira} 
\ead{arvieira@fisica.ufmg.br}
\author[PT]{Brigitte Hiller}
\ead{brigitte@teor.fis.uc.pt}
\author[PF]{A. P. Ba\^eta Scarpelli}
\ead{scarpelli.apbs@dpf.gov.br}
\author[UFMG,ING]{Marcos Sampaio}
\ead{marcos.sampaio@durham.ac.uk}

 \address[UFMG]{Departamento de F\'{\i}sica - ICEx - Universidade Federal de Minas Gerais\\ P.O. BOX 702, 30.161-970, Belo Horizonte - MG - Brasil}
\address[PT]{Departamento de F\'{\i}sica, Faculdade de Ci\^encias e Tecnologia, Universidade de Coimbra, 3004-516 Coimbra - Portugal}
\address[ING]{Centre for Particle Theory, Department of Mathematical Sciences, Durham University, South Road Durham DH1 3LE, UK}
\address[PF]{Setor T\'{e}cnico-Cient\'{\i}fico - Departamento de Pol\'\i cia Federal,
Rua Hugo D'Antola, 95 - Lapa - S\~{a}o Paulo - Brazil}

\begin{abstract}
In this contribution, we present a new perspective on the control of quadratic divergences in quantum field theory, in general, and in the Higgs naturalness problem, in particular. Our discussion is essentially based on an approach where UV divergences are parameterized, after being reduced to basic divergent integrals (BDI) in one internal momentum, as functions of a cutoff and a renormalization group scale $\lambda$. We illustrate our proposal with well-known examples, such as the gluon vacuum self energy of QCD and the Higgs decay in two photons within this approach. We also discuss frameworks in effective low-energy QCD models, where quadratic divergences are indeed fundamental.
\end{abstract}

\begin{keyword}
Renormalization \sep Electromagnetic decays \sep Phenomenological quark models \sep Standard-model Higgs bosons
\end{keyword}
%

\end{frontmatter}

\section{Introduction}
\label{intro}

The Higgs boson discovery at the LHC \cite{LHC} ($m_H \approx 125\, GeV$) as well as the lack of data supporting on low energy extensions to the standard model (SM) such as supersymmetry (SuSy) has renewed the interest in possible explanations for both the electroweak hierarchy and the naturalness problem. These issues are related to a certain extent to how we interpret quadratic divergences in field theories. Two kinds of hierarchy problems arise in the SM \cite{Aoki}. The most commonly referred one, which we will simply call hierarchy problem, is related to the large radiative corrections to the Higgs mass stemming from quadratic divergences in the cutoff which supposedly cancel against the tree level value to a very high precision at the weak scale. Consequently the Higgs mass becomes quadratically sensitive to a cutoff scale $\Lambda$. On the other hand the gauge hierarchy problem has to do with logarithmic divergences which determine the running of the coupling constants: if on one level the scale that characterizes the symmetry breaking of the GUT which unifies quantum chromodynamics and electroweak theory \cite{Langacker} is $10^{14} GeV$, on the other level the electroweak symmetry breaking scale is about $10^2 GeV$. Explaining this gap is known as gauge hierarchy problem \cite{Gildener}. The solution of the hierarchy problem involves how one bypasses the quadratic divergences which, unlike other divergences of a renormalizable theory which are multiplicatively renormalized, lead to a subtractive renormalization of the Higgs boson mass. Thus the hierarchy problem  is reduced  to the naturalness of such subtraction. SuSy avoids such subtractive renormalization and would solve the technical naturalness of the SM should SuSy particles  be sufficiently light.

It is important to remark that contrarily to the interpretation  in ultraviolet (UV) complete theories, quadratic divergences cannot be excused away as an artifact of the regularization procedure by simply adopting dimensional regularization, for instance. Taking the SM as the low energy limit of a more complete theory, a cutoff must be introduced to set a landmark in which new degrees  of freedom appear. Notice that the meaning of a cutoff $\Lambda$ is twofold: it can play the role of the UV cutoff of an UV complete theory ($\Lambda \rightarrow \infty$) or a cutoff in an effective theory at which new degrees of freedom appear (merging scale). For example, for low energy models of QCD, $\Lambda \approx 1 \, GeV$, as quarks and gluons are not well defined degrees of freedom in this region \cite{Harada}. Evidently, for both the ultraviolet complete and the effective theory, naive subtraction of quadratic divergences has no effect upon the low energy dynamics. However in drawing conclusions about new physics, such subtraction becomes a subtle and relevant issue.

However, as pointed out in \cite{Masina}, the absence of quadratic divergences does not fully solve the hierarchy problem. The SM must also be UV completed at the scale $\Lambda$ by a theory without quadratic divergences. Thus the problem also passes by at which scale a complete theory (say, SuSy) appears. That is because a matching of the parameters of the high energy and low energy physics ought to guarantee light Higgs mass parameters at the merging scale. Such fine tuning would be avoided only if $\Lambda \approx 10^3 GeV$ \cite{Masina}.

Different constructions, differing by their level of sophistication, have appeared in the literature to explain away the role played by quadratic divergences in the naturalness problem, given that new physics has not been found at LHC with $\sqrt{s}= 8 \, TeV$. It is worthwhile to discuss some proposals to give a panorama on the subject. For instance, naturalness without SuSy was proposed and studied by Jack and Jones in \cite{JJ}, whereas in \cite{Kessel} it was constructed a non-SuSy hypothetical theory  which has the same particle content as softly broken minimal supersymmetric QED. It was shown that such theory was gauge invariant and free of quadratic divergences up to two loop order.

The oldest and widely discussed proposal is the Veltman condition \cite{Veltman}, by which in the SM
\be
C_V = \frac{3}{2}m_W^2 + \frac{3}{4} m_Z^2 + \frac{3}{4}m_H^2 - \sum_f n_f m_f^2,
\label{eq:VC}
\ee
$n_f=3$ for quarks and $n_f=1$ for leptons, would make the coefficient of the $\Lambda^2$ contribution to $m_H^2$ vanish if $C_V=0$. However, this leads to $m_H = 316 GeV$.

In \cite{Ma}, within the scotogenic model of neutrino masses, the introduction of two scalar doublets, distinguished by $Z_2$ symmetry, leads to two Veltman conditions  which, in principle, could satisfy the vanishing of quadratic divergences without narrowing the Higgs mass as much. An alternative to the Veltman condition would be the compositeness of the Higgs particle by a strong infrared dynamics, forming a fermionic bound state, which at high energy breaks into its elementary fermionic constituents and, hence, quadratic divergences would be absent \cite{Zubkov}.

On the grounds that neither the Veltman condition is satisfied for the measured value of the Higgs mass
(electroweak scale) nor SuSy has been found at the LHC energies, reference \cite{Masina} supposed that Veltman condition could be satisfied at some large energy $\mu_V$ where SuSy dominates (see also \cite{Chaichian}). Whereas simply imposing $C_V=0$ leads to $m_H = \approx 316 \, GeV$, at odds with the current value, in terms of physical masses and couplings (\ref{eq:VC}) is supposed to be renormalization group invariant. Thus Veltman condition (\ref{eq:VC}) at one loop order can be written in terms of running couplings as  \cite{Masina}
\be
C_V^\hbar (\mu) = 6 \lambda(\mu) + \frac{9}{4}g^2(\mu) + \frac{3}{4}g'^2(\mu) - 6 y_t^2 (\mu),
\ee
where $\mu$ is the renormalization  scale, $\lambda$ is the Higgs potential self coupling, $y_t$ is the top Yukawa coupling and $g,\,  g'$ are the electroweak gauge couplings. Setting $C_V^\hbar (\mu)=0$ allows us to infer at which scale the Veltman condition is fulfilled (higher loop order corrections leads only to a small modification in $\mu_{V_\hbar}$ \cite{Sarhi},\cite{Hamada}). A NNLO calculation for the running couplings \cite{Masina}, using as inputs $m_H = 126 \, GeV$, $\bar{m_t}(m_t) = 161.5\, GeV$ and $\alpha_3 (m_Z) = 0.1196$ for the strong coupling constant at the $Z$ boson mass, leads to $C_V^\hbar (\mu)=0$ at $\mu$ slightly larger than the Planck scale, which means that the SM is fine tuned up to this scale. Another adjustment in the parameters of the high energy fundamental theory must be performed in order to keep the Higgs and other singlets masses light at the merging scale. Such fine tuning is however unrelated to quadratic divergences. The appealing feature of this construction is that it puts off the solution of the hierarchy problem to the high energy complete UV theory.

In \cite{Craig} it was introduced new degrees of freedom through adding other contributions to Higgs boson wave function renormalization. Effectively, those new degrees of freedom  change the Higgs boson coupling and, guided by naturalness, the authors construct a weak scale effective theory in which the new extra scalar fields cancel the quadratic divergences. They also argue that the parameter space of their ``natural theories" can be tested to percent level precision through Higgs boson coupling measurements at LHC.

It is noteworthy that there have been claims which establish the Higgs lightness due to huge cancellations because an anthropic selection destroyed naturalness \cite{Agrawal}. Along similar lines, some authors claim a finite naturalness scenario in the sense that quadratic divergences are simply put aside (ignoring uncomputable power divergences) so that the Higgs mass is naturally small at least until there are no heavier particles  \cite{Farina}. They verify that finite naturalness is satisfied by the SM whilst for its extensions it remains valid only if the new physics is not much above the weak scale.

An interesting analysis on naturalness of the SM and extensions based on Bayesian statistics was performed in \cite{Fowlie}. ATLAS and CMS \cite{LHC}, \cite{CMS} operates in $20/fb$ with center of mass energy in the range $\sqrt{s}  = 7$ to $8\, TeV$ and will continue searching for SuSy to $13\, TeV$. Moreover a $\sqrt{s} = 100\, TeV$ Very Large Hadron Collider (VLHC) may be constructed \cite{Geneve}. Roughly speaking, Bayesian statistics is a numerical estimate of belief in a proposition (model), given the experimental data. Such an estimation is weighed by the Bayes-factor $B$. Unsurprisingly the evidence for the SM without quadratic divergences over SM with quadratic divergences, given both the $m_Z$ and $m_H$ measures, is huge ($B\approx 10^{30}$, given that $B=150$ is considered very strong in the Jeffrey's scale \cite{Fowlie}). A comparison between the likelihood of the SM and the constrained minimal supersymmetric SM (CMSSM) \cite{CMSSM} indicated, using as inputs the measured values of $m_H$, $m_Z$ and LHC at $20/fb$, that the Bayes factor favors the CMSSM over the SM with quadratic divergences by $\approx 10^{30}$, whereas SM without quadratic divergences is favored over the CMSSM by $\approx 700$. Before the LHC measurements, this factor would be only $\approx 2$. This is related to the ``fine tuning price". They conclude their paper arguing that  natural models are most probable and naturalness is not simply an aesthetic principle. Moreover, the fine tuning price  of null results from the VLHC ($\approx 400$) would be slightly less than that of LHC ($\approx 500$).

In this contribution, we point out another perspective on the control of quadratic divergences in quantum field theory, in general, and in the Higgs naturalness problem, in particular. Our viewpoint is consonant with the works of Fujikawa \cite{Fujikawa} and Aoki and Iso \cite{Aoki}, but justifies them at a prior regularized level led by symmetry constraints.

We illustrate our proposal with well-known examples such as the gluon vacuum self energy of QCD and the Higgs decay in two photons within this approach. We also discuss frameworks in effective low-energy QCD models, where quadratic divergences are indeed fundamental.

Our discussion is essentially based on an approach where UV divergences are parameterized, after being reduced to basic divergent integrals (BDI) in one internal momentum, as functions of a cutoff and a renormalization group scale $\lambda$ \cite{Ferreira:2011cv}-\cite{GrafenoEPL}. This construction, which was called Implicit Regularization (IR), can be generalized to arbitrary loop order to define the leading divergence of a Feynman diagram after subtraction of sub-divergences, as dictated by the local version of the BPHZ forest formula, based on the subtraction of local counter-terms \cite{Bogoliubov:1957gp}-\cite{Cherchiglia:2010yd}. Thus, it complies with locality, Lorentz invariance and unitarity. The BDI's can be absorbed in the definition of renormalization constants without being explicitly evaluated. This defines a minimal subtraction scheme, where the BDI´s are the bare bones of a Feynman amplitude UV behavior. The derivatives of BDI's with respect to an arbitrary mass scale, say $\lambda^2$, are also expressible in terms of BDI's, which, in turn, have a lower superficial degree of divergence. Therefore renormalization group functions can be consistently evaluated within this approach.

In order to address the hierarchy problem, for instance, it is necessary to introduce a cutoff.  The relations involving derivatives of BDI´s mentioned above are regularization independent and must be satisfied by any explicit regularization. Thus, we can use such relations to build a general parametrization for the BDI's in terms of $\Lambda$. Contact with other explicit regularizations such as Pauli-Villars, dimensional regularization (DReg), sharp cutoff, Proper Time, etc turns out to be immediate.

Arbitrary regularization dependent terms, which may be responsible for symmetry breaking in the underlying model, will be systematically displayed as surface terms (ST). Such ST's can be systematically derived at arbitrary loop order, being defined as specific differences between BDI's with the same superficial degree of divergence and different Lorentz structure, as we explain in the next section. The general parametrization we construct for each BDI clearly displays the ST undetermined character, which is fixed by symmetry requirements. That is because each BDI itself contains  undetermined and regularization dependent parameters. In this case, usually they can be hidden in the arbitrariness of defining a renormalization constant. However, as we shall see in the case of quadratic divergences in the hierarchy problem, they may break symmetries as well.

This work is about arbitrary regularization dependent parameters, more specifically in the case of quadratic divergences, and how symmetry constraints which fix such parameters shed light on issues such as the hierarchy problem. In the latter, the scaling argument of Bardeen \cite{Bardeen} and conformal anomaly can be used to fix a undetermined parameter in the isolated quadratic divergence which contribute to the Higgs boson mass. A generalization of this strategy to higher loops is presented. This strategy follows Jackiw proposal in \cite{JackiwFU} by which undetermined regularization dependent parameters must be fixed via symmetry and/or phenomenology constraints. In this way, we show that his strategy is accomplished not only to finite models, in which finite quantum corrections cannot be excused away by renormalization group conditions, but also to renormalizable and effective models, notably in studying quantum symmetry breakings.

To gain insight on how we deal with arbitrary parameters in a regularization independent way, as well as to give a general overview on quadratic divergences in QFT, we discuss the appearance of quadratic divergences in QCD, in the electroweak Higgs decay in two photons and in the quarkonium  light meson decay in two photons. Finally we comment on the importance of quadratic divergences in low-energy QCD effective models. We organized the presentation as follows: in section \ref{IR}, we present an overview of the Implicit Regularization approach, showing how regularization dependent terms can be consistently identified; in section \ref{QCD}, we show in the context of QCD how quadratic divergences naturally cancel themselves for the gluon self-energy; in section \ref{Higgs}, we discuss the Higgs decay to two photons, showing that the arbitrariness present in such case, although at first glance has a quadratic origin, is connected only with gauge symmetry; in section \ref{EFT}; we discuss the role played by quadratic divergences in the gap equation of Effective Field Theories of QCD; the hierarchy problem in our formalism is presented in section \ref{hie}, in which we show how it is related to the ambiguities coming from quadratic divergences; we conclude in section \ref{conc}.

\section{Basic divergent integrals, regularization ambiguities and parametrizations}
\label{IR}

In this section, we carry out a review of Implicit Regularization, as well as we discuss regularization ambiguities and general parameterizations of regularization dependent quantities. Within the approach of Implicit Regularization, the original divergent integral is assumed to be implicitly regularized (see \cite{Ferreira:2011cv}-\cite{GrafenoEPL}). This allows algebraic manipulations in the integrand. To isolate the basic loop integrals from Feynman amplitudes, the following identity,
\bea
 && \frac{1}{[(k+p)^2-m^2]} = \sum_{j=0}^{N} \frac{(-1)^j (p^2+ 2 p
\cdot k)^j}{(k^2-m^2)^{j+1}}  \nonumber \\ && +
\frac{(-1)^{N+1} (p^2 + 2 p\cdot k)^{N+1}}{(k^2 -m^2)^{N+1}
[(k+p)^2-m^2]} \, ,
\eea
can be judiciously used in the propagators, $N$ being chosen so that the external momentum dependence is extracted from the divergent loop integrals   \footnote{Such operation at the level of integrands somewhat resembles the renormalization procedure originally proposed by Bogoliubov, Parasiuk, Hepp and Zimmermann (BPHZ) \cite{Bogoliubov:1957gp}-\cite{Zimmermann:1969jj} in which divergent Green functions are Taylor expanded up to the order needed to reach convergent integrals.}. In general, besides a finite part in the UV limit, we get basic divergent integrals which in four dimensional space time are defined at one-loop as
\be
I^{\mu_1 \cdots \mu_{2n}}_{log}(m^2)\equiv \int_k \frac{k^{\mu_1}\cdots k^{\mu_{2n}}}{(k^2-m^2)^{2+n}}
\ee
and
\be
I^{\mu_1 \cdots \mu_{2n}}_{quad}(m^2)\equiv \int_k \frac{k^{\mu_1}\cdots k^{\mu_{2n}}}{(k^2-m^2)^{1+n}},
\ee
where $\int_k \equiv \int d^4k/(2 \pi)^4$. The basic divergent integrals with Lorentz indices can be expressed in terms of the ones without indices throw surface terms (ST). Such local {\emph{ regularization dependent}} surface terms are intrinsically
arbitrarily valued. Let us take one loop BDI as examples. If the integrals are $d$-dimensional, it is straightforward to show that
\bea
\Upsilon_{0}^{\mu\nu} &\equiv& \int^d_k \frac{\partial}{\partial k_{\mu}}\frac{k^{\nu}}{(k^{2}-m^{2})^{\frac{d}{2}}} \nonumber \\&=& d\Bigg[\frac{g^{\mu\nu}}{d}I_{log}(m^2)-I_{log}^{\mu\nu}(m^2)\Bigg],
\label{ST1}
\eea
\noindent
and
\bea
\Upsilon_{2}^{\mu\nu} &\equiv& \int^d_k\frac{\partial}{\partial k_{\mu}}\frac{k^{\nu}}{(k^{2}-m^{2})^{\frac{d-2}{2}}}\nonumber \\ &=& (d-2)\Bigg[\frac{g^{\mu\nu}}{(d-2)}I_{quad}(m^2)-I_{quad}^{\mu\nu}(m^2)\Bigg].
\label{ST2}
\eea
Such arbitrary surface terms are physical meaningful. The vanishing of ST's expressed by the $\Upsilon$'s reflects momentum routing invariance in the loops of a Feynman diagram \cite{Ferreira:2011cv}, \cite{Battistel:1998sz}. Spurious evaluations of such ST's are at the heart of quantum symmetry breaking by regularizations.

Next, we carry out a discussion on the regularization dependence of surface terms and BDI's. Let us start with a few four dimensional examples \cite{Varin}. Let
\be
A = \int_k \frac{k^2}{(k^2-m^2)^2},
\ee
and
\be
B = I_{quad} (m^2) + m^2 I_{log} (m^2).
\ee
We expect  $A=B$ be guaranteed by any regularization. However this is not the case. Proper-time regularization \cite{ZinnJustin:1993wc}, for instance, introduces a cutoff $\Lambda$ after Wick rotation via the following identity at the level of propagators,
\bea
\frac{\Gamma(n)}{(k^2+m^2)^n} &=& \int_0^{\infty} d\tau \tau^{n-1} e^{- \tau (k^2+m^2)} \nonumber \\ &\rightarrow& \int_{1/\Lambda^2}^{\infty}  d\tau \tau^{n-1} e^{- \tau (k^2+m^2)},
\eea
and yields
\be
A_{\Lambda}^{P.T.} = -2b (\Lambda^2 - m^2 \ln \Lambda^2/m^2),
\ee
whereas
\be
B_{\Lambda}^{P.T.} = -b (\Lambda^2 - 2 m^2 \ln \Lambda^2/m^2),
\ee
$b\equiv -i/(4 \pi)^2$. On the other hand, it is straightforward to show that standard dimensional regularization leads to $A=B$. As another example, it can be easily seen that $I_{quad}^{\mu \nu}$ evaluates differently with respect to the leading divergence in cutoff and proper-time regularization, namely
\be
I_{quad}^{\mu \nu} \stackrel{\mbox{cutoff}}{\longrightarrow} b g_{\mu \nu} \Big(-\frac{1}{4} \Lambda^2 + \frac{1}{2} m^2 \ln \frac{\Lambda^2}{m^2}\Big) + {\mbox{finite}},
\label{eqn:cutoff}
\ee
\be
I_{quad}^{\mu \nu} \stackrel{\mbox{P.T.}}{\longrightarrow} b g_{\mu \nu} \Big(-\frac{1}{2} \Lambda^2 + \frac{1}{2} m^2 \ln \frac{\Lambda^2}{m^2}\Big) + {\mbox{finite}},
\label{eqn:PT}
\ee
respectively. Before analyzing what is essentially regularization independent in these results, consider another example related to a shift in the integration variable of a four-dimensional integral, for $\omega = 2$,
\be
\Delta_1 =\int^{2 \omega}_k \frac{k_\mu}{[(k-p)^2-m^2]^2} - \int_k^{2 \omega} \frac{(k+p)_\mu}{[k^2-m^2]^2},
\label{Delta1}
\ee
in which $2 \omega$ is the dimension of the integration. Clearly $\Delta_1 =0$ in dimensional regularization because in this method no surface terms accompany shifts in the integration variable. However following Jauch and Rohrlich in \cite{Jauch:1955} one  evaluates $\Delta_1$ for $\omega$ exactly equal to 2 as
\be
\Delta_1 = \frac{-i \pi^2 (2 \pi)^4}{2} \delta_{\omega 2} p_\mu.
\ee
A similar expression may be obtained for more than linearly divergent variable shifted integrals. It is immediate from above that the Kronecker delta signs a discontinuity in the dimensionality $\omega$. An important question, given that  shifts of integration variables are regularization dependent, would be  to exploit the consequences of momentum routing invariance over regularization schemes. Some technicalities deserve attention. Symmetric integration in $n$ (integer) dimensions, namely $k_\mu k_\nu \rightarrow g_{\mu \nu} k^2/n$ under integration in $k$, for divergent integrals does $not$ hold in general. This has been a source of  disagreements in loop calculations, as discussed in \cite{PerezVictoria:2001ej} in the context of CPT violation in quantum field theory, and used in \cite{Gastmans:2011wh} to study  the Higgs decay into two photons. In particular, symmetric integration was used by  Jauch and Rohrlich in \cite{Jauch:1955} to evaluate $\Delta_1$ and it was the source of discrepancy between the results displayed in equations (\ref{eqn:cutoff}) and (\ref{eqn:PT}) as well.

We shall now construct general parameterizations for loop integrals which incorporate explicitly arbitrary regularization dependent terms which will be fixed on physical grounds. As we discussed earlier, in Implicit Regularization (IR) we display ultraviolet (and infrared\footnote{Infrared divergences can be represented by basic divergent integrals in configuration space \cite{beta}.}) divergences in terms of basic divergent integrals. Consider the regularization independent relations satisfied by the following logarithmically basic divergent integrals in $d$ (integer) dimensional spacetime:
\bea
\frac{d I_{log}(m^2)}{d m^{2}}&=&-\frac{b_{d}}{m^{2}},\nonumber\\
\frac{d I_{log}^{\mu \nu}(m^2)}{d m^{2}}&=&- \frac{g^{\mu \nu}}{d}\frac{b_{d}}{m^{2}},
\eea
where
\be
b_d = \frac{i}{(4 \pi)^{d/2}}\frac{(-)^{d/2}}{\Gamma (d/2)}.
\ee
A general parametrization involving a cutoff $\Lambda \rightarrow \infty$ that obeys the relations above is
\begin{align}
I_{log} (m^2)&=  b_{d}\ln\left(\frac{\Lambda^2}{m^2}\right) + \alpha_1,\nonumber\\
I_{log}^{\mu\nu}(m^2)&= \frac{g^{\mu\nu}}{d}\Bigg[b_{d}\ln\left(\frac{\Lambda^2}{m^2}\right) + \alpha_1'\Bigg],
\label{resilog}
\end{align}
\noindent
where $\alpha_{1}$, $\alpha'_{1}$ are arbitrary dimensionless regularization dependent constants.

\noindent
Similarly
\begin{align}
\frac{d I_{quad} (m^2)}{d m^{2}}&=\frac{(d-2)}{2}\,I_{log}(m^2),\nonumber\\
\frac{d I_{quad}^{\mu\nu}(m^2)}{d m^{2}}&= \left(\frac{d}{2}\right) I_{log}^{\mu\nu}(m^2),
\end{align}
\noindent
which leads to the general parameterizations
\small
\begin{align}
I_{quad}(m^2)=&\frac{(d-2)}{2}\Bigg[\alpha_{2}\Lambda^{2}+ b_{d}m^{2}\ln\left(\frac{\Lambda^2}{m^2}\right)+ \alpha_3 m^{2}\Bigg],\nonumber\\
I_{quad}^{\mu\nu}(m^2)=&\frac{g^{\mu\nu}}{2}\Bigg[\alpha'_{2}\Lambda^{2}+ b_{d}m^{2}\ln\left(\frac{\Lambda^2}{m^2}\right)\
+ \alpha_3' m^2\Bigg],
\label{resiquad}
\end{align}
\normalsize
in which all regularization dependence is encoded in the $\alpha$'s.

Now, if we use these parameterizations in the surface terms of equations (\ref{ST1}) and (\ref{ST2}), we get
\be
\Upsilon_0^{\mu \nu} \propto g^{\mu \nu} (\alpha_1 - \alpha_1'),
\ee
and
\be
\Upsilon_2^{\mu \nu} \propto g^{\mu \nu} [(\alpha_2 - \alpha_2')\Lambda^2 + (\alpha_3 - \alpha_3')m^2].
\ee
These results exhibit the regularization dependence of the ST. For instance, in the four-dimensional case $\Upsilon_0^{\mu \nu}=g^{\mu \nu} [i/8(4\pi)^2]$ and $\Upsilon_2^{\mu \nu}=  g^{\mu \nu}\Lambda^2 [i/4(4\pi)^2]$ in sharp cutoff regularization,  while they are both zero in DReg. As for the examples we presented earlier, it is immediate that $A=B$ within our approach because summing and subtracting $m^2$ in the numerator of $A$ leads to $B$. Whenever even powers of internal momenta appear in the numerator, one can always make use of such artifice to avoid ambiguous symmetric integration \cite{Pontes:2007fg}. For $\Delta_1$ in equation (\ref{Delta1}), one obtains
\be
\Delta_1^{IR} = \Upsilon_0^{\mu \nu} p_\nu.
\ee
As mentioned before the BDI's are the barebones of the amplitude UV behavior and can be absorbed in the definition of renormalization constants as they stand. In order to define a mass independent scheme we may trade $m^2$ for $\lambda^2 \ne 0$ and write the following regularization independent relation
\begin{equation}
I_{log}(m^{2}) = I_{log}(\lambda^{2}) + b\ln\left(\frac{\lambda^{2}}{m^{2}}\right),
\label{eqn:relog}
\end{equation}
where $\lambda$ plays the role of renormalization group scale (see \cite{Ferreira:2011cv} and references therein). However,
\bea
I_{quad}(m^2)&=&I_{quad}(\lambda^2)+ m^2 I_{log}(m^2)-\lambda^2 I_{log}(\lambda^2)\nonumber \\ &+& b(m^2-\lambda^2),
\eea
where we note that the RHS is not completely written in terms of the renormalization scale $\lambda$. That is because, as discussed in \cite{Fujikawa}, quadratic divergences, in contrast with logarithmic, must be subtractively renormalized. Renormalization group flow is essentially described by a scale
engendered by logarithmic divergences. For the sake of completeness we write out the explicit parametrization for logarithmic BDI's to arbitrary loop order. After subtraction of subdivergences according to BPHZ formalism, we may define the divergence of $n^{th}$ loop order in terms of basic divergent integrals for both massive and massless theories \cite{Cherchiglia:2010yd} in the form
\be
I_{log}^{(n)}(m^{2}) \equiv \int_{k}\frac{1}{(k^{2}-m^{2})^{2}}\ln^{n-1}\left(-\frac{(k^{2}-m^{2})}{\lambda^{2}}\right),
\ee
which obeys
\be
I_{log}^{(n+1)}(m^{2}) = I_{log}^{(n+1)}(\lambda^{2}) - b\sum_{i=1}^{n+1}\frac{n!}{i!}\ln^{i}\left(\frac{m^{2}}{\lambda^{2}}\right).
\ee
Likewise
\begin{align}
\frac{d I_{log}^{(n)}(\lambda^2)}{d
\lambda^{2}}&=-\frac{(n-1)}{\lambda^{2}}I_{log}^{(n-1)}(\lambda^2)+\frac{b_{d}}{\lambda^{2}}A^{(n)},\nonumber\\
\frac{d I_{log}^{(n)\,\mu\nu}(\lambda^2)}{d
\lambda^{2}}&=-\frac{(n-1)}{\lambda^{2}}I_{log}^{(n-1)\,\mu\nu}(\lambda^2)+\frac{g^{\mu\nu}}{2}\frac{b_{d}}{\lambda^{2}}B^{(n)}.
\label{gerder}
\end{align}
\noindent
After some algebra, one can demonstrate that the parametrization below respects (\ref{gerder})
\small
\bea
I_{log}^{(n)}(\lambda^2)&=&\sum\limits_{i=1}^{n}\frac{(n-1)!}{(i-1)!}\!\Bigg[\!\frac{(-b_{d})A^{(i)}}{(n-i+1)!}\ln^{n-i+1}\!\!\left(\frac{\Lambda^{2}}{\lambda^{2}}\right)\nonumber \\ &+& \sum\limits_{j=0}^{n-i}\frac{a_{n-j-i+1}}{j!(n-j-i)!}\ln^{j}\!\!\left(\frac{\Lambda^{2}}{\lambda^{2}}\right)\!\!\Bigg]\nonumber
\eea
and
\bea
I_{log}^{(n)\,\mu\nu}(\lambda^2)&=&\frac{g^{\mu\nu}}{2}\!\sum\limits_{i=1}^{n}\frac{(n-1)!}{(i-1)!}\!\Bigg[\!\frac{(-b_{d})B^{(i)}}{(n-i+1)!}\ln^{n-i+1}\!\!\left(\frac{\Lambda^{2}}{\lambda^{2}}\right)\nonumber \\ &+& \sum\limits_{j=0}^{n-i}\frac{a'_{n-j-i+1}}{j!(n-j-i)!}\ln^{j}\!\!\left(\frac{\Lambda^{2}}{\lambda^{2}}\right)\!\!\Bigg],
\eea
\noindent
\normalsize
where
\small
\bea
A^{(i)}&\equiv& \Gamma(d/2)\lim_{\delta\rightarrow0}\Bigg[-(i-1)\sum\limits_{l=0}^{i-2}\binom{i-2}{l}\frac{(-1)^{1+l}}{\delta^{i-2}} \nonumber \\
&\times&
\frac{\Gamma(1-\delta(i-2-l))}{\Gamma(d/2+1-\delta(i-2-l))}+\left(\frac{d}{2}\right)\sum\limits_{l=0}^{i-1}\binom{i-1}{l} \nonumber \\
&\times& \frac{(-1)^{1+l}}{\delta^{i-1}}\frac{\Gamma(1-\delta(i-1-l))}{\Gamma(d/2+1-\delta(i-1-l))}\Bigg],\nonumber\\
B^{(i)}&\equiv&\Gamma(d/2)\lim_{\delta\rightarrow0}\Bigg[-(i-1)\sum\limits_{l=0}^{i-2}\binom{i-2}{l}\frac{(-1)^{1+l}}{\delta^{i-2}} \nonumber \\
&\times& \frac{\Gamma(1-\delta(i-2-l))}{\Gamma(d/2+2-\delta(i-2-l))}+\left(\frac{d+2}{2}\right)\sum\limits_{l=0}^{i-1}\binom{i-1}{l} \nonumber \\
&\times& \frac{(-1)^{1+l}}{\delta^{i-1}}\frac{\Gamma(1-\delta(i-1-l))}{\Gamma(d/2+2-\delta(i-1-l))}\Bigg],
\eea
\noindent
\normalsize
and $a_{i}$, $a'_{i}$ are arbitrary constants. We have, for instance, the surface terms
\bea
&& \frac{1}{2}\sum_{j=1}^{n}\left(\frac{2}{d}\right)^j \frac{(n-1)!}{(n-j)!}\Upsilon_{0}^{(n)\mu\nu}=-I_{log}^{(n)\,\mu \nu}(\lambda^2)+ \nonumber \\
&&\frac{g^{\mu \nu}}{2}\sum_{j=1}^{n}\left(\frac{2}{d}\right)^j \frac{(n-1)!}{(n-j)!}I_{log}^{(l-j+1)}(\lambda^2).
\eea
\noindent
Generalization to an arbitrary number of Lorentz indices can be obtained in a similar fashion. For use in the next section, we explicitly write the ST's at one loop order up to four Lorentz indices:
\be
\Upsilon^{\mu \nu}_2 \equiv   g^{\mu \nu} I_{quad} (m^2) - 2 I_{quad}^{\mu
\nu} (m^2)  = \upsilon_1  g^{\mu \nu} \, ,
\label{CR4Q1}
\ee
\be
\Upsilon^{\mu \nu}_0 \equiv g^{\mu\nu} I_{log} (m^2) - 4 I_{log}^{\mu \nu} (m^2)
= \upsilon_2 g^{\mu \nu},
 \label{CR4L1}
\ee
\be
\Upsilon^{\mu \nu \alpha \beta}_2 \equiv
g^{\{\mu\nu}g^{\alpha\beta \}} I_{quad} (m^2)  - 8 I_{quad}^{\mu \nu \alpha
\beta}(m^2)  = \upsilon_3 g^{\{\mu\nu}g^{\alpha\beta \}} \, ,
\label{CR4Q2}
\ee
\be
\Upsilon^{\mu \nu \alpha \beta}_0 \equiv
g^{\{\mu\nu}g^{\alpha\beta \}}  I_{log} (m^2)  - 24   I_{log}^{\mu \nu \alpha
\beta} (m^2)  =  \upsilon_4 g^{\{\mu\nu}g^{\alpha\beta \}}.
\label{CR4L2}
\ee
where $\upsilon$'s are arbitrary regularization dependent constants and the curly brackets stand for symmetrization.

\section{Example: Cancelation of Quadratic Divergences and Renormalization of QCD at one loop}
\label{QCD}

In this section, we show that quadratic divergences that appear in gluon self energies cancel out as they should, since they organize themselves into quadratic surface terms which are set to zero on gauge invariance grounds. We take the opportunity to evaluate the beta function of QCD using a different approach from the one presented in \cite{Ottoni}. In the present case, we will show, relying on the parameterization of BDI's just presented in the last section, how a cutoff can be introduced while respecting gauge invariance.

For completeness, we present the bare QCD Lagrangian,
\begin{eqnarray}
{\cal{L}}_0 &=& \frac{1}{4}(F_{ 0 \mu \nu }^{a})^{2}-\frac{1}{2\alpha }
(\partial ^{\mu }A_{0\mu }^{a})^{2} \nonumber \\ &+&
\bar{\psi}_{0}^{i} (i \gamma_\mu D_\mu^{ij} - m_0 \delta^{ij}) \psi_0^j  + \\
&+& i (\partial^\mu \bar{c}_0^a)D_\mu^{ab}c_0^b,
\label{eqn:lag}
\end{eqnarray}
and the definition of the counterterms in function of the renormalization constants
\be
A_{0 \mu}^a = Z_3^{1/2} A_\mu^a \,\,\, , \,\,\, c_0^a = \tilde{Z}_3^{1/2} c^a
\,\,\, , \,\,\, \psi_0 = Z_2^{1/2} \psi \,\,\, ,
\nonumber
\ee
\be
g_0 = Z_g g \,\,\, , \,\,\, m_0 = Z_m  m \, .
\label{eqn:bare}
\ee
Thus, ${\cal{L}}_0 = {\cal{L}} + {\cal{L}}_{ct}$, where
${\cal{L}}$ is  equal to  ${\cal{L}}_0$,  except that it is written in
terms of the renormalized variables, whereas    ${\cal{L}}_{ct}$ is the
counterterm Lagrangian, which reads
\bea
{\cal{L}}_{ct} &=& (Z_3-1)\frac{1}{2} A_a^\mu \delta^{ab} (g_{\mu \nu}
\partial^2 - \partial_\mu \partial_\nu)A_b^\nu  + \nonumber \\
&+& (\tilde{Z}_3-1)\bar{c}^a\delta_{ab}(-i \partial^2) c^b  \nonumber \\ &+&
(Z_2 - 1)\bar{\psi}^i (i \gamma^\mu \partial_\mu - m) \psi^i  \nonumber \\
&-& (Z_2 Z_m -1) m \bar{\psi}^i \psi^i \nonumber \\  &-& (Z_1-1) \frac{1}{2} g
f^{abc} (\partial_\mu A^a_\nu - \partial_\nu A^a_\mu) A_b^\mu A_c^\nu
\nonumber \\ &-& (Z_4-1) \frac{1}{4} g^2 f^{abe}f^{cde} A^a_\mu A^b_\nu A_
c^\mu A_ d^\nu \nonumber \\ &-& (\tilde{Z}_1 -1) i g f^{abc} (\partial^\mu
\bar{c}^a) c^b A^c_\mu \nonumber \\ &+& (Z_{1F} -1)g\bar{\psi}^i t^a_{ij}
\gamma^\mu \psi^j A^a_\mu  \, ,
\eea
where we have defined
$$
Z_1 \equiv Z_g Z_3^{3/2} \,\,\, , \,\,\, Z_4 \equiv Z_g^2 Z_3^2 \,\,\, ,
$$
$$
\tilde{Z}_1 \equiv Z_g \tilde{Z}_3 Z_3^{1/2} \,\,\, , \,\,\, Z_{1F} \equiv Z_g
Z_2 Z_3^{1/2} \, .
$$
The equality of $Z_g$ for all the couplings leads to the Slavnov-Taylor
identities:
\be
\frac{Z_1}{Z_3}=\frac{\tilde{Z}_1}{\tilde{Z}_3}=\frac{Z_{1F}}{Z_2}=\frac{Z_4}{Z_
1} \,  .
\label{eqn:sti}
\ee
The Feynman rules for QCD can be found in any textbook. We work in the Feynman gauge, where $\alpha = 1$. We will focus mainly on the gluon self-energy and the three-gluon vertex, from which the beta function at one loop order can be computed. Further details can be found in \cite{Ottoni}.
\begin{figure}[!h]
\begin{center}
 \includegraphics[scale=0.5]{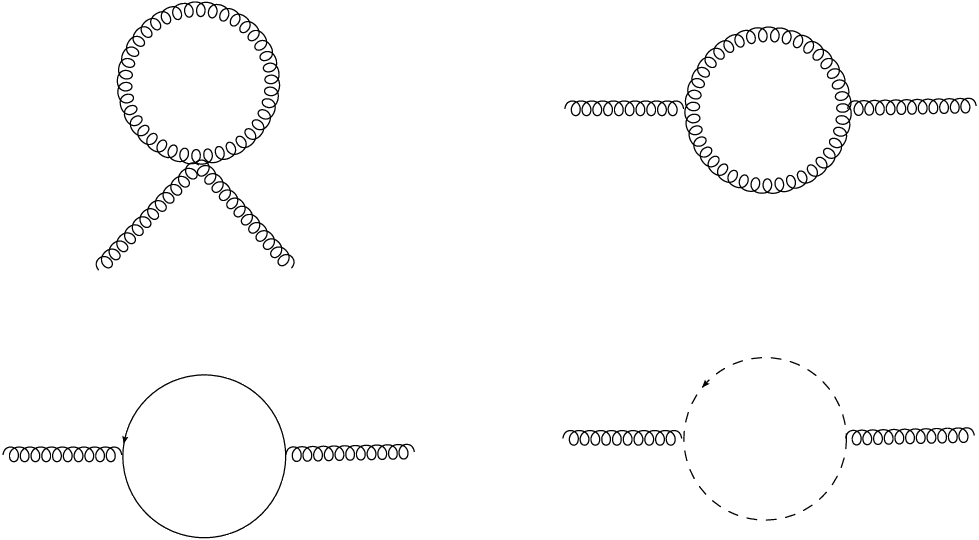}
\end{center}
\caption{One-loop gluon self-energy}
\label{gluonself}
 \end{figure}

The gluon self-energy is composed of  four
contributions, as depicted in figure \ref{gluonself},
\begin{equation}
\Pi _{\mu \nu }^{ab}=\Pi _{\mu \nu }^{ab}(1)+\Pi _{\mu \nu }^{ab}(2)+\Pi
_{\mu \nu }^{ab}(3)+\Pi _{\mu \nu }^{ab}(4),  \label{pimini}
\end{equation}
where $\Pi _{\mu \nu }^{ab}(1)$, $\Pi _{\mu \nu }^{ab}(2)$, $\Pi _{\mu \nu
}^{ab}(3)$ and $\Pi _{\mu \nu }^{ab}(4)$ represent the gluon tadpole, the
gluon loop, the ghost loop and the quark loop, respectively.
It is purely transversal as required by the Slavnov-Taylor identities and thus it does not admit a mass
term and there should be no mass renormalization. Hence, the quadratic
divergences which appear in    $\Pi _{\mu \nu }^{ab}$ should cancel out.

We begin with the gluon tadpole,
\begin{eqnarray}
\Pi^{ab}_{\mu \nu}(1) &=& -g^2C_2(G) \delta^{ab}3\int_k \frac{g_{\mu \nu}}
{k^2-\mu^2}\nonumber \\
&=&    -3 g^2 g_{\mu \nu} C_2(G) \delta^{ab} I_{quad}(\mu^2),
\end{eqnarray}
in which $\mu$ is fictitious mass which should be set to zero in the end. At this point, one may argue that $I_{quad}(\mu^2) =0$ as $\mu \rightarrow 0$, but in an general calculation this may not be the case. We shall carry the quadratic divergences until the end, so that all regularization dependent parameters are ultimately fixed by symmetry. This approach is adequate for interpreting the role of quadratic divergences in the examples we exploit in the next sections.

The gluon loop amplitude reads
\begin{equation}
\Pi^{ab}_{\mu \nu}(2)= \frac{(-i)^2}{2} \int_k  g^2 f^{acd} f^{bcd} N_{\mu \nu}
\frac{1}{k^2-\mu^2}\frac{1}{(k+p)^2-\mu^2} \,\, ,
\label{eqn:Pmn2}
\end{equation}
where
\begin{eqnarray}
N_{\mu \nu}&=& 2p_\mu p_\nu -5(p_\mu k_\nu +p_\nu k_\mu) -10 k_\mu k_\nu \nonumber \\
&-& g_{\mu \nu}[(p-k)^2+ (k+2p)^2] \, ,
\end{eqnarray}
which yields
\begin{eqnarray}
\Pi^{ab}_{\mu \nu}(2)&=& -\frac 12 g^2 C_2(G) \delta^{ab} [(2p_\mu p_\nu -
4p^2 g_{\mu \nu})J (p^2,\mu^2)  \nonumber \\
&-& g_{\mu \nu}(2 I_{quad}(\mu^2) + p^\alpha p^\beta
\Upsilon_{\alpha \beta}^0)   \nonumber \\
&-& 10(\, p_\nu J_\mu (p^2,\mu^2) + J_{\mu \nu}(p^2,\mu^2) \, )] \,  .
\end{eqnarray}
As for  the ghost loop, we have
\begin{eqnarray}
\Pi^{ab}_{\mu \nu}(3)&=& - g^2f^{dac}f^{cbd}  \int_k \frac{i^2}{k^2-\mu^2}
\frac { (p+k)_\mu k_\nu}{[(k+p)^2-\mu^2]}\nonumber \\
&=& - g^2 \delta^{ab} C_2(G) (p_\nu J_\mu  (p^2,\mu^2) \nonumber \\ &+& J_{\mu
\nu} (p^2,\mu^2) ).
\end{eqnarray}
The integrals $J_{\mu \nu}$, $J_\mu$ and $J$ are given by
\bq
&& J_{\mu \nu} (p^2, \mu^2\rightarrow 0) = I_{quad\, \mu \nu} (\mu^2)-
 p^2 I_{log\, \mu \nu} (\mu^2) \nonumber \\
&& + 4 p^\alpha p^\beta  I_{log\, \mu \nu \alpha \beta}
+ b \Bigg\{ \frac {p_\mu p_\nu}{3} \Big[\frac 16 - \ln \Big( \frac{-p^2}{e^2 \mu^2}\Big) \Big] \nonumber \\
&&- \frac{p^2 g_{\mu \nu}}{6}\Big[\frac
13 - \frac{1}{2} \ln \Big(\frac{-p^2}{e^2 \mu^2}\Big)\Big] \Bigg\} \, , \\
&& J_\mu (p^2,\mu^2 \rightarrow 0) = -2p^\alpha I_{log\, \alpha \mu} (\mu^2) \nonumber \\
&&+ \frac b2 \, p_\mu  \ln \Big(\frac{-p^2}{e^2 \mu^2}\Big)
\eq
and
\be
J (p^2,\mu^2 \rightarrow 0) = I_{log}(\mu^2) -b \ln \Big(\frac{-p^2}{e^2 \mu^2}\Big).
\ee

The fermion loop contribution to the gluon self energy is identical to the
vacuum  polarization tensor of $QED$, except for the colour and number of
fermions ($n_f$) factors. It has been computed within IR \cite{PRD2} and reads
\begin{eqnarray}
&&\Pi_{\mu \nu}^{ab}(4) = \frac{4}{3} g^2  C(r) n_f
\delta^{ab}   \Bigg\{       \Big(p_\mu p_\nu-p^2 g_{\mu \nu} \Big)  \times
\nonumber \\ && \times
\Bigg[ I_{log} (\mu^2) - b  \Bigg( \ln \Big(-\frac{p^2}{e^2 \mu^2} \Big)
+ \frac{1}{3} \Bigg) \Bigg]  \nonumber \\
&& + \Upsilon^2_{\mu \nu}(\mu^2) + p^2 \Upsilon^{(0)}_{\mu \nu} +
p^\alpha p^\beta \Upsilon^{(0)}_{\mu \nu \alpha \beta} + \nonumber \\
&&p^\alpha p_\mu \Upsilon^{(0)}_{\nu \alpha} + p^\beta
p_\nu \Upsilon^{(0)}_{\mu \beta} + p^\alpha p^\beta  g_{\mu
\nu} \Upsilon^{(0)}_{\alpha \beta}  \Bigg\} \, .
\end{eqnarray}
Altogether, $\Pi _{\mu \nu }^{ab} = \sum_{i=1}^4 \Pi_{\mu \nu}^{ab}(i)$ reads
\begin{eqnarray}
&& \Pi _{\mu \nu }^{ab}(p^2, \lambda^2) = -\frac{b}{9} g^{2}(p^{2}g_{\mu \nu
}-p_{\mu }p_{\nu })\delta ^{ab}  \notag \\ &&\times \Bigg\{
i \Big[ \frac{5}{3}C_{2}(G)-\frac{4}{3}n_{f}C(r) \Big] I_{\log }(\lambda ^{2})
\nonumber \\ && +\Big( 15C(r)-6n_{f} \Big) \ln \Big(\frac{\lambda
^{2}}{p^{2}}\Big)-2C (r)+2 n_{f} \Bigg\} + \nonumber \\&& \Bigg\{
\Upsilon^2_{\mu \nu}(\lambda^2)-\lambda^2\Upsilon^{(0)}_{\mu \nu} + p^2 \Upsilon^{(0)}_{\mu \nu}+
p^\alpha p^\beta \Upsilon^{(0)}_{\mu \nu \alpha \beta} + \nonumber \\
&&p^\alpha p_\mu \Upsilon^{(0)}_{\nu \alpha} + p^\beta
p_\nu \Upsilon^{(0)}_{\mu \beta} + p^\alpha p^\beta  g_{\mu
\nu} \Upsilon^{(0)}_{\alpha \beta}\Bigg\} \nonumber \\
&& \times g^2 \delta^{ab} \Big( C_2(G) + \frac 43 C(r) n_f\Big),
\label{eqn:Pmnf}
\end{eqnarray}
where we used relation (\ref{eqn:relog})  in order to introduce a renormalization group scale $\lambda$. Notice that the infrared divergences, as $\mu \rightarrow 0$, cancel out as they should and only the quadratic surface term has a dependence on $\lambda^2$ (we use, for simplicity $\Upsilon^{(0)}_{\alpha \beta}(\mu^2)\equiv \Upsilon^{(0)}_{\alpha \beta}$). We also have used the relation
\be
\Upsilon^{(2)}_{\alpha \beta}(\mu^2)=\Upsilon^{(2)}_{\alpha \beta}(\lambda^2 \ne 0)+(\mu^2-\lambda^2)\Upsilon^{(0)}_{\alpha \beta}.
\ee
Setting the surface terms to zero or, accordingly, making $\upsilon's = 0$ in (\ref{CR4Q1})-(\ref{CR4L2}) renders the total amplitude transverse, as required by gauge invariance. This amounts to exercising a constrained gauge invariant version of IR (CIR) \cite{Ferreira:2011cv}. Notice that the quadratic divergences organized themselves as surface terms. If evaluated in dimensional regularization, they yield zero because the latter is a gauge invariant framework. We shall systematically set the surface terms to zero and express the BDI's as a function of $\lambda$ until the end of this section.

We define the counterterm for the amplitude (\ref{eqn:Pmnf}) by minimally subtracting the BDI expressed by $I_{log} (\lambda^2)$:
\begin{equation}
Z_{3}=1-i\left[ \frac{5}{3}C_{2}(G)-\frac{4}{3}n_{f}C(r)\right] I_{\log
}(\lambda ^{2})g^{2}+O(g^{3}).
\label{z3}
\end{equation}

\begin{figure}[!h]
\begin{center}
 \includegraphics[scale=0.4]{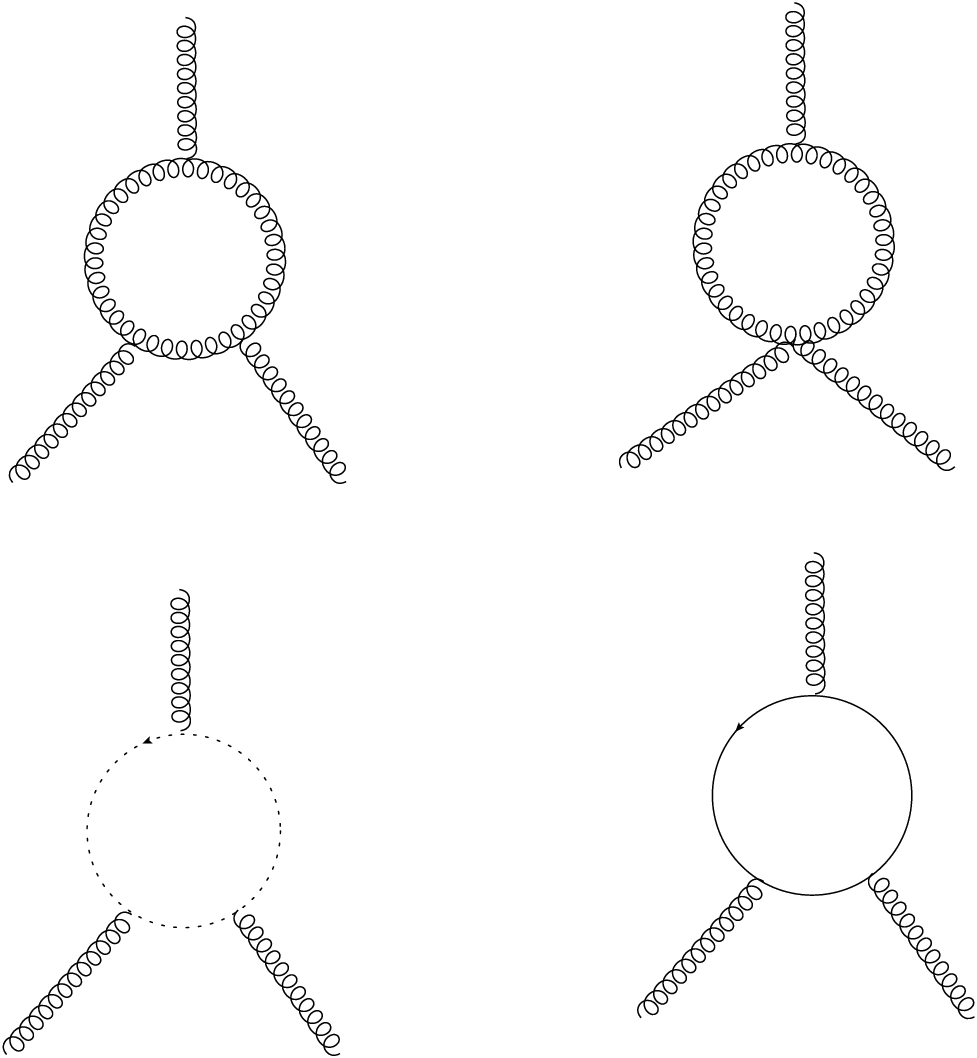}
\end{center}
\caption{One-loop three-gluon vertex}
\label{3gluon}
 \end{figure}

The class of one loop three-gluon vertex graphs, from which we shall define
$Z_ 1$, are shown in fig. \ref{3gluon}.

For the sake of brevity, we shall present only the result here. Let $p$ and $q$ be the external momenta. Then
\bea
&& \Lambda^{abc}_{\mu \nu \lambda}(p,q) = -i g f^{abc}V_{\mu \nu
\lambda}(p,q,p+q)  \nonumber \\ && \times \Bigg(- i g^2 \Big( -\frac{2}{3}
C_2(G) + \frac{4}{3} C(r) n_f \Big) I_{log} (\lambda^2) +  Z_1 -
1 \Bigg)\nonumber \\ && + \widetilde{\Lambda}^{abc}_{\mu \nu \lambda}(p,q)      \, ,
\eea
$V_{\mu \nu \lambda}(p,q,p+q) = (p-q)_\lambda g_{\mu \nu} -p_{\mu} g_{\nu
\lambda} + q_{\nu} g_{\mu \lambda} $,  from which we define
\be
Z_1 = 1 + i g^2 \Bigg(- \frac{2}{3} C_2(G) + \frac{4}{3} C(r) n_f
\Bigg)I_{log}(\lambda^2) \, .
\ee

Since we are mainly interested in the computation of the beta function, we will not present the other renormalization constants. However, as explicitly showed in \cite{Ottoni}, all of them obey Slavnov-Taylor identities expressed by (\ref{eqn:sti}):
\begin{align}
\frac{Z_1}{Z_3}=\frac{\tilde{Z}_1}{\tilde{Z}_3}=\frac{Z_{1F}}{Z_2}=&\frac{Z_4}{Z_
1} = 1 + i g^2 C_2(G) I_{log}(\lambda^2)\nonumber\\
=& 1 + i g^2 C_2(G)\Bigg[b\ln\left(\frac{\Lambda^2}{\lambda^2}\right) + \alpha_1\Bigg].
\label{eqn:stir}
\end{align}
In the last line, we make use of the parameterization of BDI's introduced in the last section. Since the Slavnov-Taylor identities are manifestations of gauge invariance, we are intimately showing how a cutoff can be introduced without breaking gauge symmetry. In other words, after we disentangle BDI's and surface terms, we can safely introduce a cutoff in the former by using the parameterizations, since all symmetry breaking terms are encoded in the latter.

To conclude this section, we compute the beta function of QCD at one-loop level. As explained before, we have introduced the parameter $\lambda$, which will play the role of renormalization group scale. Thus, we have
\be
\beta (g) = \lambda \frac{\partial g}{\partial \lambda}.
\ee
Recall (\ref{eqn:bare}): $g_0 = Z_g g$, $Z_g = Z_1 Z_3^{-3/2}$. Hence,
\be
2 \lambda^2 \frac{\partial}{\partial \lambda^2} \Big( Z_g g \Big) =0
\Longrightarrow \beta (g) = - 2 g \lambda^2 \frac{\partial \ln Z_g}{\partial \lambda^2} \, .
\ee
Now, using that
\be
\lambda^2 \frac{\partial}{\partial \lambda^2} I_{log} (\lambda^2) = - b
\ee
in the equation above yields, after some simple algebra,
\be
\beta = - \frac{g^3}{3 (4 \pi)^2} \Big( 11 C_2(G) - 4 C(r) n_f \Big) + O(g^5) \,
.
\ee
In a similar fashion, we can work with an explicit cutoff, in which case the renormalization group scale would be $\Lambda$, on the basis of a cutoff parameter independence of the Green's functions in the Wilsonian renormalization group \cite{Wilson}. In this case,  the renormalization constants will explicitly depend on $\Lambda$, for instance,
\be
Z_1 = 1 + i g^2 \Bigg(- \frac{2}{3} C_2(G) + \frac{4}{3} C(r) n_f
\Bigg)\Bigg[b\ln\left(\frac{\Lambda^2}{\lambda^2}\right) + \alpha_1\Bigg],
\ee
which furnishes the following result
\be
\lambda^2 \frac{\partial }{\partial \lambda^2}\ln Z_g(g,\Lambda^2/\lambda^2)=-\Lambda^2 \frac{\partial }{\partial \Lambda^2}\ln Z_g(g,\Lambda^2/\lambda^2).
\ee
Thus, after simple algebra, we obtain the same result for the beta function.

In summary, an explicit cutoff can always be introduced once we had correctly identified the regularization dependent terms. The surface terms would boil down to the arbitrary terms in the parameterization of $I_{log}$'s and $I_{quad}$'s and would be fixed by gauge symmetry as well. The advantage of working with basic divergent integrals is that we can neatly identify regularization dependent terms as surface terms. Finally, this procedure can be worked out to general loop order \cite{David}, \cite{Edson},\cite{Adriano1}.

\section{Higgs decay to two photons}
\label{Higgs}

\begin{figure}[h]
\begin{center}
\includegraphics[scale=0.45]{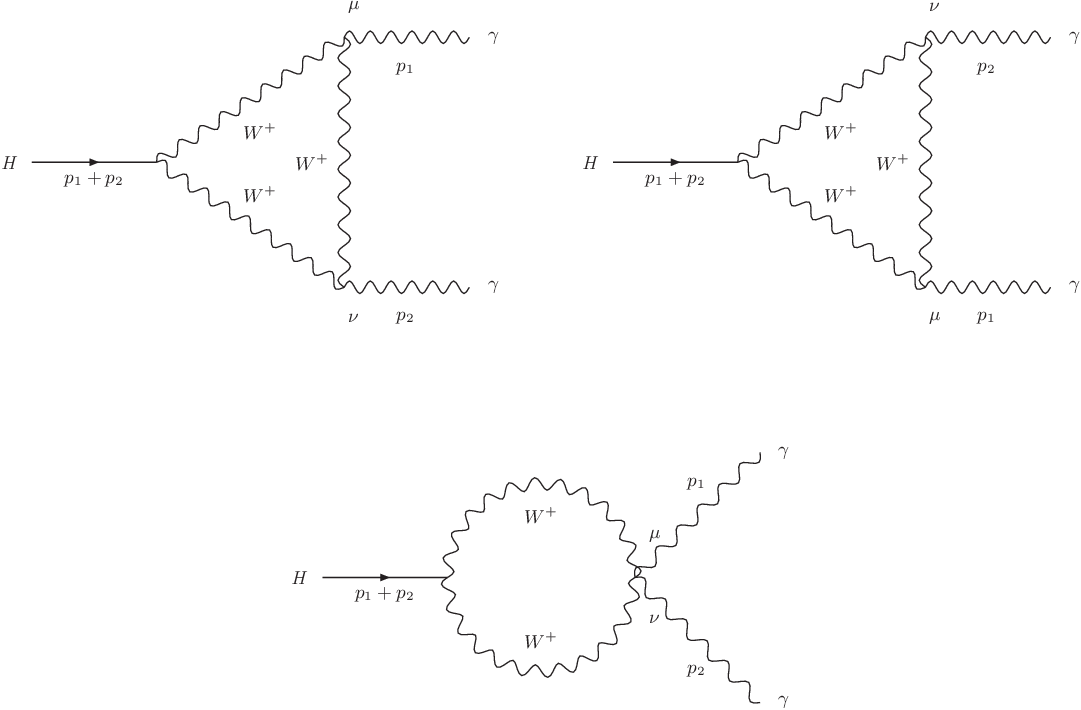}
\end{center}
\vspace{-0.5cm}
\caption{Diagrams that contribute to the Higgs decay to two photons}
\label{diagrams}
\end{figure}

In this section we will discuss how to fix arbitrariness involved in the calculation of the Higgs decay to two photons. This decay was a subject of discussion in the recent literature (\cite{Adriano2} and references therein). We follow the framework presented in section \ref{IR}. We will consider only the W boson loop, since it already contains all relevant aspects regarding arbitrariness we intend to discuss in the following. As showed in \cite{Adriano2}, the diagrams in the unitary gauge that contribute are shown in figure \ref{diagrams}.

\noindent
The contributions can be simplified to
\bq
&&M=ie^{2}g M_{w}\Big[M_{\mu\nu}^{(a)}+M_{\mu\nu}^{(b)}+M_{\mu\nu}^{(c)}\Big]({\epsilon_{1}}^{\mu})^{\ast}({\epsilon_{2}}^{\nu})^{\ast} \nonumber\\
&&+(p_{1}\leftrightarrow p_{2},\mu\leftrightarrow\nu),
\eq
with
\bq
&&M_{\mu\nu}^{(a)}=-\frac{4}{M^{2}_{w}}\Big[g_{\mu\nu}(p_{1})^{\alpha} (p_{2})^{\beta}I^{(3)}_{\alpha\beta}+(p_{1}\cdot p_{2})I^{(3)}_{\mu\nu}\nonumber\\
&&-(p_{1})_{\nu}(p_{2})^{\alpha}I^{(3)}_{\mu\alpha}-(p_{2})_{\mu}(p_{1})^{\alpha}I^{(3)}_{\nu\alpha}\Big]+\nonumber\\
&&+\frac{2}{M^{2}_{w}}\Big[g_{\mu\nu}(p_{1}\cdot p_{2})-(p_{2})_{\mu}(p_{1})_{\nu}\Big]I^{(3)}_{2},\label{Mabef}\\
&&M_{\mu\nu}^{(b)}=\int\limits_{k}\frac{3(g_{\mu\nu}k^{2}-4k_{\mu}k_{\nu})}{(q_{1}^{2}-M^{2}_{w})(q_{2}^{2}-M^{2}_{w})(q_{3}^{2}-M^{2}_{w})},\label{Mbbef}\\
&&M_{\mu\nu}^{(c)}=6g_{\mu\nu}\Big[(p_{1}\cdot p_{2})I^{(3)}_{0}-(p_{1})^{\alpha}I^{(3)}_{\alpha}-\frac{M^{2}_{w}}{2}I^{(3)}_{0}\Big]\nonumber\\
&&+6\Big[2(p_{1})_{\nu}I^{(3)}_{\mu}-(p_{2})_{\mu}(p_{1})_{\nu}I^{(3)}_{0}\Big],\label{Mcbef}\\
\label{Before}
&&I^{(3)}_{0,2,\mu,\mu\nu}=\int\limits_{k}\frac{1,k^2,k_{\mu},k_{\mu}k_{\nu}}{(q_{1}^{2}-M^{2}_{w})(q_{2}^{2}-M^{2}_{w})(q_{3}^{2}-M^{2}_{w})}.
\eq

As showed in \cite{Adriano2}, there is an inherent arbitrariness in the expressions above which will present itself as a surface term in our framework. Explicitly, we have\footnote{We define $\tau=\frac{M^{2}_{h}}{4M^{2}_{w}}$ and \begin{equation}
f(\tau)=\left\{\begin{array}{lcc}
\arcsin^2(\sqrt{\tau})&\mbox{for}&\tau\leq 1\,,\\[5mm]
-{\displaystyle\frac{1}{4}\,\left[\ln\frac{1+\sqrt{1-\tau^{-1}}}{1-\sqrt{1-\tau^{-1}}}-i\pi\right]^2}
&\mbox{for}&\tau>1\,.\end{array}\right.\nonumber
\end{equation}}

\begin{align}
M_{\mu\nu}^{(a)}=\frac{\big[(p_{2})_{\mu}(p_{1})_{\nu}-g_{\mu\nu}(p_{1}\cdot p_{2})\big]}{M^{2}_{w}}\Bigg[\frac{i}{16\pi^2}-2\upsilon_{2}\Bigg],
\label{Ma}
\end{align}
\begin{align}
M_{\mu\nu}^{(b)}+M_{\mu\nu}^{(c)}&=\frac{i}{16\pi^2M^{2}_{w}}\big[(p_{2})_{\mu}(p_{1})_{\nu}-g_{\mu\nu}(p_{1}\cdot p_{2})\big]\times\nonumber\\&\times\Bigg[\frac{3\tau^{-1}}{2}+\frac{3(2\tau^{-1}-\tau^{-2})f(\tau)}{2}\Bigg]\nonumber\\
&+g_{\mu\nu}(p_{1}\cdot p_{2})\Bigg(\frac{3\tau^{-1}}{2M^{2}_{w}}\upsilon_{2}\Bigg).
\label{Mb}
\end{align}

As one can immediately notice, the surface term in the second expression breaks gauge invariance. Therefore, as in the last section, the imposition of such symmetry will fix the ambiguity to a precise value (in the present case, it will be null), furnishing the well-established value for this decay.

Now we would like to perform the same analysis again, but from a different point of view, which will clarify the role played by gauge symmetry and quadratic divergences.

We begin by defining the amplitude
\bq
&&M_{\mu\nu}=ie^{2}gM_{w}\Big[M_{\mu\nu}^{(a)}+M_{\mu\nu}^{(b)}+M_{\mu\nu}^{(c)}\Big] \nonumber \\
&&+(p_{1}\leftrightarrow p_{2},\mu\leftrightarrow\nu),
\eq
\noindent
which, if gauge invariance is to be guaranteed, must satisfy
\begin{align}
M_{\mu\nu}p_{1}^{\mu}p_{2}^{\nu}=0.
\end{align}

By using expressions (\ref{Mabef}), (\ref{Mbbef}) and (\ref{Mcbef}) one obtains
\bq
&&M_{\mu\nu}p_{1}^{\mu}p_{2}^{\nu}=ie^{2}gM_{w}\Bigg[\int\limits_{k}\frac{3}{(k-p_{1})^{2}-M^{2}_{w}}-\int\limits_{k}\frac{3}{k^{2}-M^{2}_{w}}\nonumber\\
&&+\int\limits_{k}\frac{3}{(k-p_{2})^{2}-M^{2}_{w}}-\int\limits_{k}\frac{3}{(k-p_{1}-p_{2})^{2}-M^{2}_{w}}\Bigg]
\label{main}
\eq

This is the main result of this section. Firstly, we notice the appearance of quadratic divergent integrals which could indicate that the arbitrariness stemming in the Higgs decay to two photons shares a common origin with the hierarchy problem. However, this is not the case. To demonstrate this, one may resort to the general parametrization of quadratic divergences presented in eq. (\ref{resiquad}). As can be easily seen, there is an arbitrary parameter ($\alpha_{2}$) multiplying a cutoff $\Lambda^2$ which is in the root of the hierarchy problem as will be explained in the next section. For the present case, however, this arbitrariness will play no role, since we have a difference between quadratic divergent integrals, which results in the cancelation of such coefficient. Therefore, the ambiguity in the present case must have a different origin.

Secondly, we notice that the expression above can be related to the following tadpole
\begin{figure}[h]
\begin{center}
\includegraphics[scale=0.45]{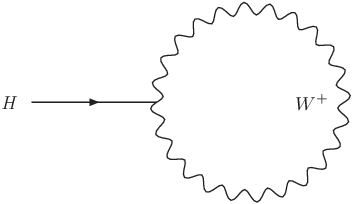}
\end{center}
\vspace{-0.5cm}
\caption{Tadpole}
\label{Tadpole}
\end{figure}

\noindent
whose analytical expression is given by
\be
T=-\frac{g}{M_{w}}\int\limits_{k}\Bigg[1-\frac{3M^{2}_{w}}{k^{2}-M^{2}_{w}}\Bigg].
\ee
The first term is a quartic divergent integral which can, in principle, be added and subtracted to expression (\ref{main}) in order to reproduce the tadpole exactly. However, since such integrals do not depend on any physical scale, they are completely unphysical and should be discarded. Therefore, one can immediately notice that the result (\ref{main}) is a difference between tadpoles with different momentum routing. As explained in \cite{Ferreira:2011cv}, the condition to implement momentum routing is just to
demand that the difference between the same Feynman diagram with different momentum routing is null. Therefore, one can easily notice that the condition to have a gauge invariant result for the Higgs decay is just to demand momentum routing
invariance of the tadpole depicted in fig \ref{Tadpole}. In other words, by demanding this tadpole to be momentum routing invariant, gauge symmetry will be automatically respected.

To conclude, we will demonstrate that the gauge breaking term is exactly the same we had before in eq. (\ref{Mb}). We begin using the following expansion
\be
f(k+a)=f(k)+a_{\sigma}\frac{\partial}{\partial k_{\sigma}}f(k)+\frac{a_{\sigma}a_{\rho}}{2!}\frac{\partial^{2}}{\partial k_{\sigma}k_{\rho}}f(k)+\cdots,
\ee
which in our case is given by
\begin{align}
\frac{1}{(k+a)^2-M^{2}_{w}}=&\frac{1}{k^2-M^{2}_{w}}-2a_{\sigma}\frac{k_{\sigma}}{(k^{2}-M^{2}_{w})^{2}}\nonumber\\&-a_{\sigma}a_{\rho}\frac{\partial}{\partial k_{\rho}}\frac{k_{\sigma}}{(k^{2}-M^{2}_{w})^{2}}+\cdots.
\end{align}
Thus, following the reasoning of \cite{Ferreira:2011cv}
\begin{align}
\int\limits_{k}\frac{1}{(k+a)^2-M^{2}_{w}}-\int\limits_{k}\frac{1}{k^2-M^{2}_{w}}=-a^{2}\upsilon_{2}.
\end{align}
Replacing this result into eq. (\ref{main}), we finally obtain
\begin{align}
M_{\mu\nu}p_{1}^{\mu}p_{2}^{\nu}&=ie^{2}gM_{w}6(p_{1}\cdot p_{2})\upsilon_{2}\nonumber\\
&=ie^{2}gM_{w}(p_{1}\cdot p_{2})^{2}\Bigg(\frac{3\tau^{-1}}{M^{2}_{w}}\upsilon_{2}\Bigg)
\end{align}
which is exactly what the imposition of the Ward identity to eq. (\ref{Mb}) would furnish\footnote{There is a extra two factor due to the sum of the crossed diagram.}.

In summary, we clarified, by using the second approach, that the Higgs decay to two photons is not connected to quadratic ambiguities.
The arbitrariness inherent of this calculation comes from the subtraction of logarithmic divergent integrals and is fixed by gauge invariance. We would like also to stress that, since in our framework regularization dependent terms can be consistently identified, it is possible to introduce a cutoff without breaking gauge invariance. In other words, after the identification of the surface terms (which control the symmetry breaking), any divergent amplitude will be written in terms of BDI's which, by means of the general parametrizations presented in section \ref{IR}, will depend explicitly on a cutoff $\Lambda^2$.

\section{Quadratic Divergences and Effective Theories: Nambu-Jona-Lasinio model}
\label{EFT}

Contrarily to the previous examples, quadratic divergences play a vital role in the description of dynamical chiral symmetry breaking in models of low energy QCD, such as the Nambu--Jona-Lasinio model \cite{Nambu:1961}. This model belongs to the class of non-renormalizable Lagrangians and the regulator, usually expressable in terms of a cutoff $\Lambda$ for the UV divergent one-loop quark integrals appearing at leading order of $N_c$, is characteristic of the scale at which spontaneous breakdown of chiral symmetry occurs, typically of the order of 1 GeV. We illustrate the role of the quadratic divergence in terms of  the original 2 flavor NJL model applied to the light quarks with $N_c=3$
\begin{eqnarray}
\label{LNJL}
{\cal L}_{NJL}&=&{\bar \psi}(x) (i {\gamma^\mu \partial_\mu} -m_c ) \psi(x) \nonumber \\
&+& \frac{G}{2}[({\bar \psi}(x) \psi(x))^2 +({\bar \psi}(x) i\gamma_5 {\tau_i} \psi(x))^2],
\end{eqnarray}
with $G>0$. There are two relevant quantities needed to be considered to set the scale of chiral symmetry breaking, one related to the weak decay constant of the pion $f_\pi \sim 93$ MeV, which is a logarithmically divergent integral and the other with the gap equation, which is quadratically divergent. We will consider for simplicity the chiral limit, $m_c=0$. The integrals are regulated using a Pauli-Villars regularization \cite{Pauli-Villars} with two subtractions in a form which is equivalent to the sharp Euclidean cutoff when scalar integrals are considered \cite{Osipov:1985}. Other regularizations have been discussed in \cite{Harada}, all displaying a quadratic divergence for the gap equation, and leading to similar conclusions. For comparison we will also use the general parametrization of eqs. (18) and (21). While $f_\pi$ depends explicitly only on the constituent quark mass $M$ and $\Lambda$,
\begin{eqnarray}
\label{fpi}
f_\pi^2&=&-4 N_c M^2 i I_{log}^\Lambda (M^2) \nonumber \\
&=& \frac{N_c M^2}{(2 \pi)^2} \Big(\ln(1+\tilde{\lambda}^2)-\frac{\tilde{\lambda}^2}{1+\tilde{\lambda}^2}\Big)
\end{eqnarray}
where $\tilde \lambda^2=\frac{\Lambda^2}{M^2}$, the gap equation also depends on the coupling strength of the four quark interaction $G$,
\begin{equation}
\label{gap}
M-m_c =  M\frac{N_c G M^2}{(2 \pi)^2}(\tilde \lambda^2-ln(1+\tilde \lambda^2))
\end{equation}
which has to reach a critical value $G_{cr}$ for the phase transition from the Wigner-Weyl phase ($M=0$ in the chiral limit) to the asymmetric phase to occur.

The solution of the gap equation corresponds to the minimum of the effective potential  $V(\sigma,\pi)$ calculated to lowest order in $N_c$ counting, with $\sigma={\bar \psi}(x)\psi(x), \pi_i={\bar \psi}(x)i \gamma_5 \tau_i\psi(x)$ the standard auxiliary bosonic variables (see e.g. \cite{Osipov:2000})
\begin{eqnarray}
\label{Veff}
&&V(\sigma,\pi_i)= \frac{\sigma^2 +\pi_i^2}{2 G} \Big(1-\frac{N_c G \Lambda^2}{4 \pi^2}\Big) \nonumber \\
&&+ \frac{N_c}{8 \pi^2} \Big[(\sigma^2 +\pi_i^2)^2 \ln\Big(1+\frac{\Lambda^2}{\sigma^2 +\pi_i^2}\Big) \nonumber \\
&& -\Lambda^4 \ln \Big(1+\frac{\sigma^2 +\pi_i^2}{\Lambda^2}\Big)\Big]
\end{eqnarray}

Taking the expectation value in the vacuum $<\pi_i>=0$, the minimum  is localized at $<\sigma>=M\ne 0$, if the curvature $$C=\partial^2_\sigma V(\sigma,\pi_i)_{|\sigma=0,\pi_i=0} < 0 $$ corresponding to the onset of a mexican hat shaped potential. The latter condition leads to $G> G_{cr}=\frac{2 \pi^2}{N_c \Lambda^2}$.

This critical value is however constrained by the empirical value of $f_\pi$, which sets a minimal value for the cutoff $\Lambda_{cr}\sim .72$ GeV, obtained by solving eq. \ref{fpi}. This result has been shown long ago in \cite{Schaden:1988}: below this critical value there are no solutions for the given value of $f_\pi$ and above this value two branches emerge as functions of ($M,\Lambda$), representing asymptotically a strong coupling regime (branch 1) where the quark mass goes faster to infinity than the cutoff, $\tilde \lambda^2=\frac{\Lambda^2}{M^2}\rightarrow 0$, and a weak coupling regime (branch 2) with $\tilde \lambda^2=\frac{\Lambda^2}{M^2}\rightarrow \infty$. Both regimes are in the phase of spontaneous breakdown of chiral symmetry, but empirical values of the light constituent quark masses $200\prec M\prec 400$ MeV rule out branch 1.  
Branch 2 is characterized by a coupling $N_c G \Lambda^{2} \sim {\cal O}(1)$. This is the result of the gap equation with the $f_\pi$ constraint and evidences a quadratic divergence, while for branch 1 one would obtain asymptotically $Gf_{\pi}^2 \sim {\cal O}(1)$, \cite{Schaden:1988}.

One can thus infer that the quadratic divergence of the gap equation is necessary to ensure a sensible solution for the values of the constituent quark masses together with the empirical value of $f_\pi$ in the phase of spontaneously broken chiral symmetry, at leading order of $N_c$ counting. Furthermore the condition $\Lambda \ge \Lambda_{cr}$ must be fulfilled, whereby $\Lambda_{cr}$ is uniquely determined by the logarithmic divergence associated with $f_\pi$.

The meson mass spectrum, in this case the Goldstone pion and the $\sigma$-meson with $m_\sigma= 2 M$, emerge as consequence of dynamical chiral symmetry breaking with the cutoff dependence completely absorbed in the value of the constituent quark mass.

Further light can be shed on the relevance of the quadratic divergence: in a recent extended version of the NJL model, which contemplates the most general combinations of spin zero multiquark interactions relevant at the scale of chiral symmetry breaking, including a complete set of explicit symmetry breaking interactions \cite{Osipov1:2013},\cite{Osipov2:2013}, it is shown that $\Lambda$ associated with the quadratic divergences of the gap equation can be used to establish a counting scheme which allows to classify all relevant interactions  (i.e. which survive in the limit of $\Lambda\rightarrow \infty$) in the phase of spontaneous symmetry breaking. This counting scheme is in consonance with the large $N_c$ counting scheme and requires the $\Lambda^2$ behavior of the gap equation.

If instead one would use the parameterizations (18) and (21) for the logarithmic and quadratic divergence one would obtain
\begin{equation}
\label{fpiIR}
f_\pi^2=\frac{N_c M^2}{(2 \pi)^2} (\ln(\tilde{\lambda}^2)-\alpha_1)
\end{equation}
and for the gap equation
\begin{equation}
\label{gapIR}
M-m_c =  M\frac{N_c G M^2}{(2 \pi^2)}(\alpha_2\tilde{\lambda}^2 -\ln(\tilde{\lambda}^2)+\alpha_3)
\end{equation}
which for $\tilde{\lambda}^2\gg 1$, $\alpha_2=\alpha_1=1,\alpha_3=0$ reduces to the result using Pauli-Villars regularization.

Knowing that the curvature of the effective potential is in this case given by
$$C=\partial^2_\sigma V(\sigma,\pi_i)_{|\sigma=0,\pi_i=0} =\frac{1}{2 G}-\frac{N_c \alpha_2 \Lambda^2}{4 \pi^2},$$ the absence of the quadratic divergence in the gap equation, obtained by choosing $\alpha_2=0$, leads to $C=\frac{1}{2 G}$; since $G>0$ one deduces immediately that only the symmetric phase is described in this case, independently of any parameters of the model. The implicit regularization thus corroborates the fact that the presence of quadratic divergences is essential to be able to reach the phase of spontaneously broken chiral symmetry.

\section{Hierarchy problem}
\label{hie}

As discussed in the introduction, many proposals have been devised to interpret the role and fate of quadratic divergences in the hierarchy problem.
It is necessary to introduce a cutoff to serve as a merging scale when we study the SM as an effective theory. A general parametrization for ultraviolet divergences with an explicit scale $\Lambda$ much greater than the characteristic masses of the model can be constructed, as we have demonstrated in earlier sections. The parametrization of quadratic divergence embodies a regularization dependent coefficient multiplying $\Lambda^2$. As a result of negative searches for SuSy at the TeV scale, its original motivation of solving the hierarchy problem by canceling out the quadratic divergences becomes questionable. This is because the stability at quantum level of the hierarchy EW scale $\ll M_{Planck}$ becomes more difficult to respect, although some extensions in MSSM have been envisaged \cite{Antoniadis}.  The natural question, following the examples we presented in previous sessions, is whether or not it can be fixed on symmetry grounds \cite{JackiwFU}.

Because of its chiral nature, the Lagrangian of the standard model possesses conformal invariance, except for the Higgs mass term, which is related to the hierarchy problem. Bardeen \cite{Bardeen} has argued that, once the classical conformal invariance and its minimal violation by quantum anomalies are imposed on the SM, it can be freed from quadratic divergences (and hence the hierarchy problem) and one can, in principle, directly interpolate the electroweak scale and the Planck scale. Such idea has been taken forward to envisage extensions to the SM with a flat Higgs potential at the Planck scale \cite{Orikasa}.

It remains to establish this hypothesis into  a calculational framework. This is exactly where the general parametrization for basic divergent integrals is useful. We use the Bardeen's hypothesis as a symmetry guide to fix an arbitrary parameter multiplying $\Lambda^2$ to zero.  This has also an aesthetic appeal, since we would be left with logarithmic divergences which can be multiplicatively renormalized.

Weinberg was the first to examine the complications caused by quadratic divergences in a mass independent renormalization scheme \cite{Weinberg}. Fujikawa in \cite{Fujikawa} addressed this problem by introducing a counterterm independent of the scalar mass to subtract the quadratic divergent contribution. This was done in a similar fashion as Callan avoided the quadratic divergence by a mass insertion technique in his Callan-Symanzik equation \cite{Callan}. This in turn is closely related to the scaling argument of Bardeen \cite{Bardeen} \footnote{ As we will see, this is exactly what dimensional
regularization does in terms of subtracting the quadratic divergence, that is setting the arbitrary parameter  multiplying $\Lambda^2$ to zero.}.

Moreover, as explained by Aoki and Iso in \cite{Aoki},
at classical level the Higgs mass term breaks scale invariance of the SM, which would have an increase of symmetry should the mass term vanish. Such increase of symmetry has no role in controlling divergences, since scale invariance is broken by the logarithmic runnings of the couplings. Once quadratic divergences are subtracted, the trace of the energy momentum tensor becomes proportional to $\Delta m^2 H^\dagger H + \beta_{g_i} {\cal{O}}_i$, in which $\Delta m^2 \propto m^2$ and not $\Lambda^2$. The anomalous term and the mass term are soft breaking terms, since they do not generate quadratic divergences. Still according to \cite{Aoki}, in the Wilsonian renormalization group, quadratic divergences determine a position of the critical surface of the theory, and the scaling behavior around such critical surface is determined by logarithmic divergences. The subtraction of quadratic divergences, according to \cite{Aoki}, then amounts to a coordinate transformation in the theory space and, thus, such divergences have no role whatsoever in the fine tuning problem. As we will see, these conclusions can be reached already at regularized level using the symmetry argument proposed in \cite{Bardeen} to fix arbitrary regularization dependent parameters.

Consider the Higgs sector of the SM  Lagrangian,
\begin{eqnarray}
\mathcal{L}_H(x)&=&[D^{\mu}\Phi(x)]^{\dagger}[D_{\mu}\Phi(x)]-\mu^2\Phi(x)^{\dagger}\Phi(x)
\nonumber \\
&-&\lambda[\Phi(x)^{\dagger}\Phi(x)]^2,
\label{eq13}
\end{eqnarray}
where $\Phi(x)=\begin{pmatrix}\phi^+\\\phi^0\end{pmatrix}$ is the Higgs field, $D_{\mu}$ is the covariant derivative that couples it with the gauge fields and $\mu^2<0$. The scale transformations
\be
x'=e^{-\alpha}x
\label{eq14a}
\ee
and
\be
\phi'(x)= e^{-\alpha d}\phi(e^{-\alpha}x)
\label{eq14}
\ee
leave (\ref{eq13}) unchanged for $\mu^2=0$, where $\alpha$ here is a scale parameter and $d$ is the scale dimension of the field. The mass term breaks that classical conservation law because it is the only one in the Lagrangian which does not possess scale dimension equal to four, i. e. the mass  did not transform according to the rules (\ref{eq14a}) and (\ref{eq14}). Therefore,
\begin{equation}
\Theta^{\mu}_{\mu}=m^2\Phi(x)^{\dagger}\Phi(x),
\label{eq15}
\end{equation}
where $m^2=-2\mu^2$ is the tree level Higgs mass.

Since quantum corrections make the couplings depend on  the renormalization group scale, the Lagrangian (\ref{eq13}) changes due to the scale transformation
\be
x'=e^{-\alpha}x \rightarrow \Lambda'=e^{\alpha}\Lambda,
\ee
which leads to
\bea
&& \delta\mathcal{L}_H(m_H(\Lambda),\lambda(\Lambda))= \nonumber \\
&& \alpha\{m^2_H(\Lambda) \gamma \Phi(x)^{\dagger}\Phi(x)+\beta_{\lambda}[\Phi(x)^{\dagger}\Phi(x)]^2\},
\label{eq16}
\eea
where $m^2_H(\Lambda)$ and $\lambda(\Lambda)$ are the renormalized Higgs mass and self-coupling, respectively, and
\bea
 \gamma=\frac{\Lambda^2}{m_H^2(\Lambda^2)}\frac{\partial m_H^2(\Lambda^2)}{\partial \Lambda^2}
\label{eq17}
\eea
is the renormalization group gamma function. Hence, the complete violation of the dilatation current, due to the mass term and quantum corrections, is given by

\begin{equation}
\Theta^{\mu}_{\mu}=(m^2+m^2_H(\Lambda) \gamma) \Phi(x)^{\dagger}\Phi(x)-\beta_{\lambda}[\Phi(x)^{\dagger}\Phi(x)]^2.
\label{eq18}
\end{equation}
We may now write, for the Higgs renormalized mass, according to equation (\ref{resiquad}) \cite{Vieira:2012ex},
\bea
m^2_H(\Lambda) &=& m^2- \frac{3 \alpha_{2}}{8 \pi^2 \upsilon^2}[m_{Z}^2+2m_{W}^2+m^2-4m_{t}^2]\Lambda^2\nonumber\\
	&+& O\left(\ln\frac{\Lambda}{m}\right).
\label{eq19}
\eea
Using (\ref{eq19}) in (\ref{eq17}), we get
\begin{equation}
m^2_H(\Lambda)\gamma=- \frac{3 \alpha_{2}}{8 \pi^2 \upsilon^2}[m_{Z}^2+2m_{W}^2+m^2-4m_{t}^2]\Lambda ^2+O(m^2).
\label{eq20}
\end{equation}
Now if we try to restore the classical limit, taking  $m\rightarrow 0$ and $\beta_{\lambda}\rightarrow 0$ in equation (\ref{eq18}), the
only term that spoils the recovery of the dilatation current conservation is
\bea
\centering
  \Theta_{\mu}^{\mu}= \frac{-3 \alpha_{2}}{8 \pi^2 \upsilon^2}[m_{Z}^2+2m_{W}^2-4m_{t}^2]\Lambda ^2\Phi(x)^{\dagger}\Phi(x).
  \label{eq21}
\eea
In a non-supersymmetric scenario in order to restore the classical limit ($\Theta_{\mu}^{\mu}=0$), we have to choose $\alpha_2=0$.

%

Let us consider some numerical implications of our results. We can estimate in which scale the fine-tuning starts and perturbation
theory breaks by asking where  $\left|\delta m^2\right|= O(m^2)$. Considering $\alpha_2=-1$ (obtained in a sharp cutoff regularization),
the experimental data for the masses ($m_{t}=173 GeV$, $m_{W}= 80.2 GeV$, $m_{Z}= 91.2 GeV$ and $m=126\ GeV$) and the VEV value
($\upsilon=246\ GeV$) that scale would be $\Lambda\approx 0.5\ TeV$. It means that the SM model as an effective theory should be reliable
up to this scale and new physics should appear beyond it. However, as mentioned before, this new physics has not been found with
$\sqrt{s}=8\ TeV$.

Nevertheless, we can choose $\alpha_2=0$ considering the arguments above, which leave us with the logarithmic correction to the Higgs
mass given by \cite{Vieira:2012ex},
\be
 \delta m ^ 2 = \frac{3 m^ 2}{16 \pi^2 \upsilon^2}[2m_{t}^2+2m_{W}^2+m^2-m_{Z}^2] \ln \frac {\Lambda^ 2}{m^2_H}.
 \label{22}
\ee

The estimate for the SM cutoff now is extremely large. We have $\left|\delta m^2\right|= O(m^2)$ when $\Lambda\approx
10^{7}\ TeV$. We conclude that consistency of scale symmetry breaking avoids the fine-tuning and makes perturbation theory
($\left|\delta m^2\right|<< O(m^2)$) reliable up to this scale.

We end this section making connection with dimensional regularization. Let us write basic quadratic divergent integral $I_{quad}(m^2)$
\begin{equation}
I_{quad}(m^2)= \lim_{\mu\rightarrow 0}\int_k\frac{1}{(k^2-m^2-\mu^2)}.
\label{eq2}
\end{equation}
We can write (\ref{eq2}) as
\bea
&& \int_k\frac{1}{(k^2-m^2-\mu^2)}=I_{quad}(\mu^2)+m^2 I_{log}(\mu^2)+\nonumber\\
&&+m^4\int_k\frac{1}{(k^2-\mu^2)^2(k^2-m^2-\mu^2)}.
\label{eq3}
\eea
Using (\ref{eqn:relog}) and
\begin{align}
&\int_k \frac{1}{(k^2-\mu^2)^2(k^2-m^2-\mu^2)}= \nonumber\\
&- b\left(1+\frac{\mu^2}{m^2}\right)\ln\left[\frac{m^2}{\mu^2}\left(1+\frac{\mu^2}{m^2}\right)\right] +\frac{b}{m^2},
\label{eq5}
\end{align}
we obtain
\begin{equation}
I_{quad}(m^2)= \lim_{\mu\rightarrow 0}I_{quad}(\mu^2)+m^2I_{log}(m^2)+\frac{i}{(4\pi)^2} m^2.
\label{eq6}
\end{equation}
In dimensional regularization it is well known that
\begin{equation}
\lim_{\mu\rightarrow 0} I_{quad}(\mu^2)=0,
\label{eq7}
\end{equation}
which ultimately implies that $\alpha_2 =0$, taking our general parameterizations for $I_{quad}$ and $I_{log}$ into account.

As a final comment, we would like to emphasize that the procedure adopted in this paper is applicable to higher order calculations. Should we take our parametrization to two loop order, it is not difficult to show that \cite{Hamada}
\be
\int_{k}\int_{q} \frac{1}{k^2 q^2 (k+q)^2} = a_1 \Lambda^2
+ a_2 \mu^2 +a_3 \mu^2 \ln \frac{\Lambda^2}{\mu^2},
\ee
where $\mu$ is a mass infrared regulator and $a_1$, $a_2$ and $a_3$ are arbitrary finite constants, which are combinations of constants of integrations. The equation above is an example of a general parametrization of a typical leading quadratic divergence of two-loop order. As we can see, it is possible in our approach to adjust arbitrary finite constants in higher order calculations so as to have a null contribution from quadratic divergences.

\section{Concluding remarks}
\label{conc}

In this paper a discussion was carried out on the role of quadratic divergences in quantum field theory. This discussion was based in a general parameterization of basic divergent integrals. These basic divergent integrals, obtained in the context of Implicit Regularization, contain all the divergent content of a given amplitude. The parameterization we adopted embodies possible results coming from different regularization procedures. Arbitrary constants which naturally appear in the procedure can be adjusted so as to enforce symmetries of the model or experimental results.

We present some examples in which the cancellation of quadratic divergences plays a pivotal role, namely in the Higgs decay to two photons and in the one-loop renormalization of QCD. In both cases we exemplify with our formalism how one can introduce a cutoff without breaking gauge symmetry. We also present an example in effective field theories in which the presence of quadratic divergences is fundamental to obtain phenomenological meaningful results. Finally we discuss the hierarchy problem. It is shown that the classical scaling argument of Bardeen and conformal anomaly can be used as a symmetry guide to fix the arbitrary regularization dependent parameter in the isolated quadratic divergence which contributes to the Higgs mass.

\section*{Acknowledgments}

M.S., A.R.V and A.L.C. thank CNPq and FAPEMIG for financial support. M.S. thanks Durham University for the kind hospitality. This work has been partially supported by the Funda\c{c}\~ao para a Ci\^encia e
Tecnologia, the iniciative QREN,
financed by UE/FEDER through COMPETE - Programa Operacional Factores de
Competitividade. This research is part of the EU Research Infrastructure
Integrating Activity Study of Strongly Interacting Matter (HadronPhysics3)
under the 7th Framework Programme of EU, Grant Agreement No. 283286.

This work is dedicated to the memory of Prof. Maria Carolina Nemes.

\end{document}